\documentclass[twocolumn]{aastex631}

\usepackage[T1]{fontenc}
\usepackage[utf8]{inputenc}
\usepackage{CJKutf8}
\usepackage{xcolor}
\usepackage{graphicx}
\usepackage{listings}
\usepackage{amsmath}
\usepackage{comment}
\usepackage{tabularx}
\usepackage{booktabs}
\usepackage{ragged2e}
\definecolor{mauve}{rgb}{0.88, 0.69, 1.0}
\definecolor{oldmauve}{rgb}{0.4, 0.19, 0.28}
\begin{document}

\title{Variations in the Milky Way's Stellar Mass Function at [Fe/H] < -1}

\correspondingauthor{Jiadong Li}
\email{jdli@mpia.de}

\author[0000-0002-3651-5482]{Jiadong Li}
\affiliation{Max-Planck-Institut für Astronomie, Königstuhl 17, D-69117 Heidelberg, Germany}

\author[0000-0003-4996-9069]{Hans-Walter Rix}
\affiliation{Max-Planck-Institut für Astronomie, Königstuhl 17, D-69117 Heidelberg, Germany}

\author[0000-0001-5082-9536]{Yuan-Sen Ting}
\affiliation{Department of Astronomy, The Ohio State University, Columbus, USA}
\affiliation{Center for Cosmology and AstroParticle Physics (CCAPP), The Ohio State University, Columbus, OH 43210, USA}
\affiliation{Max-Planck-Institut für Astronomie, Königstuhl 17, D-69117 Heidelberg, Germany}

\author[0009-0007-8574-0890]{Yu-Ting Wang}
\affiliation{Key Laboratory of Space Astronomy and Technology, National Astronomical Observatories, CAS, Beijing 100101, China}
\affiliation{Max-Planck-Institut für Astronomie, Königstuhl 17, D-69117 Heidelberg, Germany}
\affiliation{University of Chinese Academy of Sciences,  Beijing 100049, China}

\author{Szabolcs~M{\'e}sz{\'a}ros}
\affiliation{ELTE E\"otv\"os Lor\'and University, Gothard Astrophysical Observatory, 9700 Szombathely, Szent Imre H. st. 112, Hungary}
\affiliation{MTA-ELTE Lend{\"u}let ``Momentum" Milky Way Research Group, 9700 Szombathely, Szent Imre H. st. 112, Hungary}
\affiliation{HUN-REN CSFK, Konkoly Observatory, Konkoly Thege Mikl\'os \'ut 15-17, Budapest, 1121, Hungary}

\author[0000-0003-3410-5794]{Ilija Medan} 
\affiliation{Department of Physics and Astronomy, Vanderbilt University, Nashville, TN 37235, USA}

\author[0000-0002-1802-6917]{Chao Liu}
\affiliation{Key Laboratory of Space Astronomy and Technology, National Astronomical Observatories, CAS, Beijing 100101, China}
\affiliation{University of Chinese Academy of Sciences,  Beijing 100049, China}

\author[0000-0001-7395-1198]{Zhiqiang Yan}
\affiliation{School of Astronomy and Space Science, Nanjing University, Nanjing, 210000, China}
\affiliation{Key Laboratory of Modern Astronomy and Astrophysics, Nanjing University, Ministry of Education, Nanjing 210093, China}

\author[0000-0002-7489-5244]{Peter J. Smith}
\affiliation{Max-Planck-Institut für Astronomie, Königstuhl 17, D-69117 Heidelberg, Germany}
\affil{Fakultät für Physik und Astronomie, Universität Heidelberg, Im Neuenheimer Feld 226, 69120 Heidelberg, Germany}

\author[0000-0002-8280-4808]{Dan Qiu}
\affiliation{Key Laboratory of Space Astronomy and Technology, National Astronomical Observatories, CAS, Beijing 100101, China}

\author[0000-0002-1379-4204]{Alexandre Roman-Lopes}
\affiliation{Department of Astronomy, Universidad de La Serena, Av. Raul Bitran \#1302, La Serena, Chile}

\author[0000-0001-5417-2260]{Gregory M. Green}
\affiliation{Max-Planck-Institut für Astronomie, Königstuhl 17, D-69117 Heidelberg, Germany}

\author{Danny Horta}
\affiliation{Institute for Astronomy, University of Edinburgh, Royal Observatory, Blackford Hill, Edinburgh, EH9 3HJ, UK}

\author[0000-0003-0179-9662]{Zachary Way}
\affiliation{Department of Physics and Astronomy, Georgia State University, 25 Park Place, Atlanta, GA 30303, USA}

\author{Tadafumi Matsuno}
\affiliation{Astronomisches Rechen-Institut, Zentrum f\"ur Astronomie der Universit\"at Heidelberg, M\"onchhofstra{\ss}e 12-14, 69120 Heidelberg, Germany}

\author{Stefano O. Souza}
\affiliation{Max-Planck-Institut für Astronomie, Königstuhl 17, D-69117 Heidelberg, Germany}

\author{Jos\'e G. Fern\'andez-Trincado}
\affiliation{Universidad Cat\'olica del Norte, N\'ucleo UCN en Arqueolog\'ia Gal\'actica - Inst. de Astronom\'ia, Av. Angamos 0610, Antofagasta, Chile}
\affiliation{Universidad Cat\'olica del Norte, Departamento de Ingenier\'ia de Sistemas y Computaci\'on, Av. Angamos 0610, Antofagasta, Chile}

\begin{abstract}
We present the first determination of the Galactic stellar mass function (MF) for low-mass stars ($0.2-0.5~{\rm M}_\odot$) at metallicities [Fe/H]$\lesssim -1$. A sample of $\sim$53,000  stars was selected as metal-poor on the basis of both their halo-like orbits and their spectroscopic [Fe/H] from Gaia DR3 BP/RP (XP) spectra. These metallicity estimates for low-mass stars were enabled by calibrating Gaia XP spectra with stellar parameters from SDSS-V. 
For $-1.5<$[Fe/H]$<$-1, we find that the MF below 0.5 ${\rm M}_\odot$ exhibits a ``bottom-heavy'' power-law slope of $\alpha \sim -1.6$. We tentatively find that at even lower metallicities, the MF becomes very bottom-light, with a near-flat power-law slope of $\alpha \sim 0$  that implies a severe deficit of low-mass stars. This metallicity-dependent variation is insensitive to the adopted stellar evolution model. These results show that the Galactic low-mass MF is not universal, with variations in the metal-poor regime.
A further calibration of XP metallicities in the regime of $M<0.5\,M_{\odot}$ and [Fe/H]$<-1.5$ will be essential to verify these tentative low-metallicity trends.
\end{abstract}

\section{Introduction} \label{sec:intro}

The stellar Initial Mass Function (IMF) is a cornerstone of astrophysics, governing the evolution of stellar populations and galaxies \citep{kroupa1993, Kroupa2002, Chabrier2003, Bastian2010}. A star's initial mass dictates its entire lifecycle and legacy, making the IMF an essential ingredient in modeling stellar populations \citep{Bastian2010, Conroy2013}, from individual star clusters to entire galaxies. This distribution of stellar masses at birth influences the chemical \citep{Yan2020}, dynamical \citep{Cappellari2012}, and photometric properties of galaxies \citep{Salpeter}.

For the bulk of stars in the Galactic disk, at $-0.4<$[Fe/H]$<0.2$, the stellar (initial) mass function seems to be well-described by Kroupa IMF\citep{kroupa1993,Kroupa2002}, which is a broken power law with a slope of $\alpha=-2.3$ for stars above $0.5~{\rm M}_\odot$, and $\alpha=-1.3$ stars below that break, referred to ``low-mass'' stars. With the Kroupa IMF serving as a canonical IMF, any stellar sub-population that has $\alpha > -1.3$ for $M<0.5~{\rm M}_\odot$ is called ``bottom-light'', any population that has $\alpha < -1.3$ is called ``bottom heavy''.

Our Milky Way's stellar halo is the closest---and hence in many ways the best---laboratory to study the formation, chemical enrichment and stellar physics of metal-poor and old stellar populations with [Fe/H]$<-1$.
The halo traces the early formation steps of our Galaxy \citep{Helmi2020}, and stars of $M_*\lesssim 0.7 {\rm M}_\odot$ play a unique role in this context: they live longer than the universe's age \citep{Hurley2000}, retain their original composition \citep{Frebel2015}, and are candidates for surviving Population III stars if they had formed at low masses \citep{Clark2011, Greif2015}.

Yet, we know little about the population of low-mass stars at [Fe/H]$\lesssim -1$, i.e. more metal-poor than the old Galactic disk. 
Recent work has shown that the stellar initial mass function (IMF) becomes increasingly bottom-light (deficient in low-mass stars) at lower ($\lesssim -0.5$) metallicities \citep{Li2023, yan2024,Qiu2025}. 
This scarcity is reflected in a long-standing discrepancy in Galactic astronomy, known as the ``M~Dwarf Problem'': the insufficient number of observed low-metallicity M~dwarfs compared to predictions from simple closed-box models of chemical evolution \citep{Woolf2012}.  
Studies have looked for M subdwarfs, empirically classified by their molecular bands \citep{Gizis1997, Lepine2007}, and \citet{Lodieu2017} assembled a spectral library ($R \sim 4$--$6\,\mathrm{k}$, $0.4$--$2.5\,\mu\mathrm{m}$) of metal-poor M dwarfs ([Fe/H] $\leq -0.5$) for direct comparison with models. 
However, even these datasets cover only moderately metal-poor objects; truly metal-poor dwarfs, say [Fe/H] $\lesssim -1.5$, remain almost unconstrained.

This scarcity is confirmed by modern surveys like SDSS/APOGEE and LAMOST, which find few low-mass stars with [Fe/H] $< -1$ \citep{Birky2020, Li2021, Qiu2025}, further conflicting with predictions from orbital dynamics \citep{Chandra2024}.  
\cite{ZhangS2019, ZhangS2021} compiled one of the largest catalogs of metal-poor M~dwarfs to date, identifying $\sim 3000$ such stars in the LAMOST survey. 
This raises a question: Do metal-poor ([Fe/H] $\lesssim -1$), low-mass stars ($M_*\lesssim 0.5 {\rm M}_\odot$) scarcely exist, or are they just hard to find? Recent radiation-hydrodynamic simulations by \citet{Bate2025} find that for star-forming clouds subjected to warm cosmic backgrounds (high redshift) or high metallicity, the resulting IMF has a deficit of brown dwarfs and low-mass stars.

The goal of this paper is to determine the stellar mass function of stars with [Fe/H]$<-1$, focusing on the low-mass regime ($M_*\lesssim 0.5 {\rm M}_\odot$). 
To construct a sample of such metal-poor stars from the vast Gaia catalog, we employ a two-step selection strategy. 
First, we use kinematic information to identify stars on halo-like (non-disk-like) orbits. This approach exploits the well-established correlation between orbital properties and metallicity \citep{Chandra2024}: stars on halo-like orbits are predominantly metal-poor ([Fe/H]$<-1$), while stars on disk-like orbits are predominantly metal-rich. 
By selecting stars with halo-like kinematics, we can efficiently remove the vast majority of metal-rich disk stars from consideration \citep[e.g][]{Lepine2007}. 
A practical issue we need to tackle is that halo-like kinematics (as proxy for kinematics of metal-poor stars) need to be identified without radial velocities that are largely unavailable for low-mass stars. 
Second, we apply a spectroscopic metallicity cut using [Fe/H] estimates derived from Gaia XP spectra that have been calibrated against SDSS-V spectra. These steps together should results a high-purity sample of low-mass stars with [Fe/H]$<-1$, whose spatial selection functioon we can characterize.

Beyond the Milky Way, observations of massive elliptical galaxies have revealed evidence for bottom-heavy IMFs, with an overabundance of low-mass stars relative to the Milky Way disk IMF \citep{vanDokkum2010, Conroy2012}. This bottom-heavy trend appears to correlate with galaxy mass and metallicity \citep{MartinNavarro2015}. Given that elliptical galaxies are thought to have formed through the hierarchical merging of smaller galaxies, similar to the process that built the Milky Way's halo, there may be a connection between the IMFs of metal-poor stars and those inferred from elliptical galaxies.

\citet{yan2024} analyze low-mass IMF variations by compiling measurements from star counting in dwarf galaxies \citep{Geha2013, Gennaro2018}, stellar population synthesis in massive early-type galaxies \citep{MartinNavarro2015, Lonoce2023}, and galactic chemical evolution modeling \citep{Yan2020}. 
Their best-fit relation reproduces the observed trend from bottom-light IMFs ($\xi_{\rm MR} \sim 0.3$--0.4) in ultrafaint dwarfs to bottom-heavy IMFs ($\xi_{\rm MR} \sim 0.6$--0.7) in metal-rich ellipticals, where $\xi_{\rm MR}$ represents the stellar mass ratio between 0.2--0.5~${\rm M}_{\odot}$ and 0.2--1~${\rm M}_{\odot}$. These discoveries present an opportunity to probe the stellar IMF in different galactic environments, potentially revealing how the IMF varies across cosmic time and in different star-forming conditions.
Studying the mass function of metal-poor stars in the Milky Way, which include both in-situ and accreted populations, offers a window into star formation processes in the early Galaxy and its satellite progenitors. Understanding the mass function at [Fe/H]$<-1$ could shed light on the M Dwarf Problem and provide constraints on models of galaxy formation and evolution.

In this paper, we measure the stellar mass function of metal-poor stars ([Fe/H] $< -1.1$) within 1~kpc of the Sun from \textit{Gaia} DR3.
As described in Section~\ref{sec:sample} we identify samples of metal-poor stars in two steps. We first use a purely kinematic selection to find stars with XP spectra whose halo-like orbits make them likely to be stars with [Fe/H]$< -1.0$. Then we ``clean" this sample by removing all stars whose XP spectra imply [Fe/H]$> -1.0$, after being calibrated to the ASPCAP scale. These two steps achieve a sample of high purity in [Fe/H]$<-1$.
In Section~\ref{section:methods}, we build a forward model that accounts for the \textit{Gaia} selection function. 
In Section~\ref{sec:results}, we find that the mass function becomes markedly bottom-light, with a turnover around $\sim0.4$--$0.6\,{\rm M}_\odot$, for metallicities below [Fe/H]$< -1.5$.

\section{Metal-Poor Star Selection}\label{sec:sample}

\begin{figure*}
    \centering
    \includegraphics[width=0.48\textwidth]{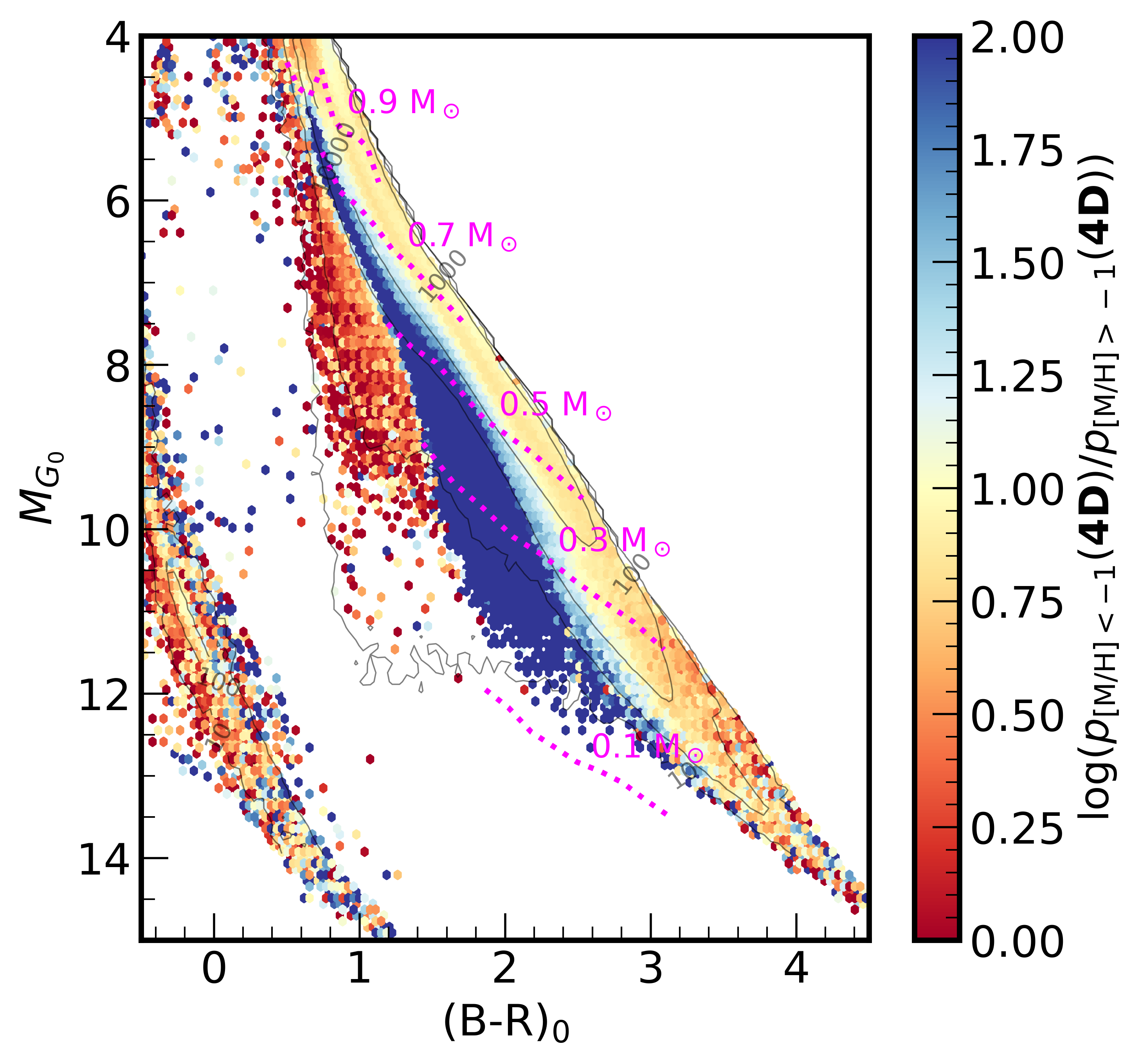}
    \includegraphics[width=0.49\linewidth]{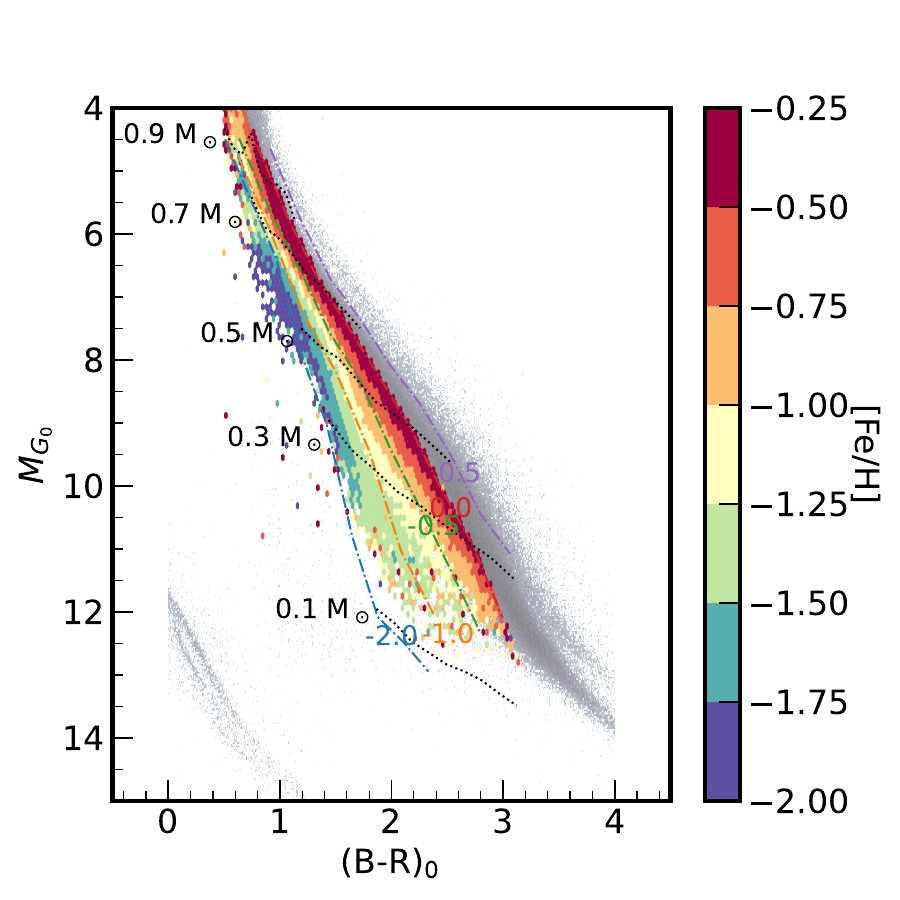}
    \caption{Color-Magnitude Diagrams (CMDs) of the Gaia XP sample. 
    \textit{Left:}  The CMD of Gaia sources within 1 kpc is color-coded by $\eta$, defined in Eq.~\ref{eq:eta}, which is the likelihood that a star's orbit is halo-like rather than disk-like. Blue colors denote stars that are more likely to be on halo-like orbits.
    \textit{Right:} CMD of the sample with halo-like orbits ($\eta>10$), color-coded by XP-derived metallicity. 
    Typical absolute-magnitude error is dominated by distance uncertainties, with $\sigma_{M_G}\approx 0.08~\mathrm{mag}$ at $d=800~\mathrm{pc}$.
    The gray background represents sources within 100 pc. 
    The dot--dashed colored lines show loci from the PARSEC stellar evolution models evaluated at a fixed age of 5~Gyr for different metallicities. The black dotted lines indicate iso-mass lines from the PARSEC models.
    }
    \label{fig:cmd_eta_feh}
\end{figure*}

\begin{figure*}
    \centering
    \includegraphics[width=0.49\linewidth]{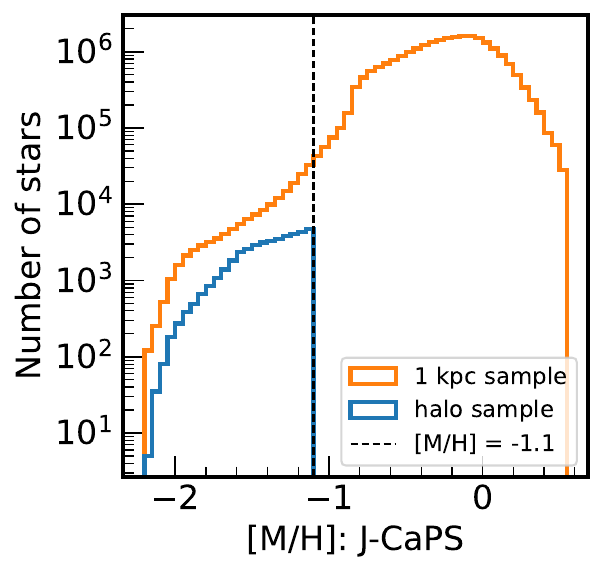}
    \includegraphics[width=0.48\linewidth]{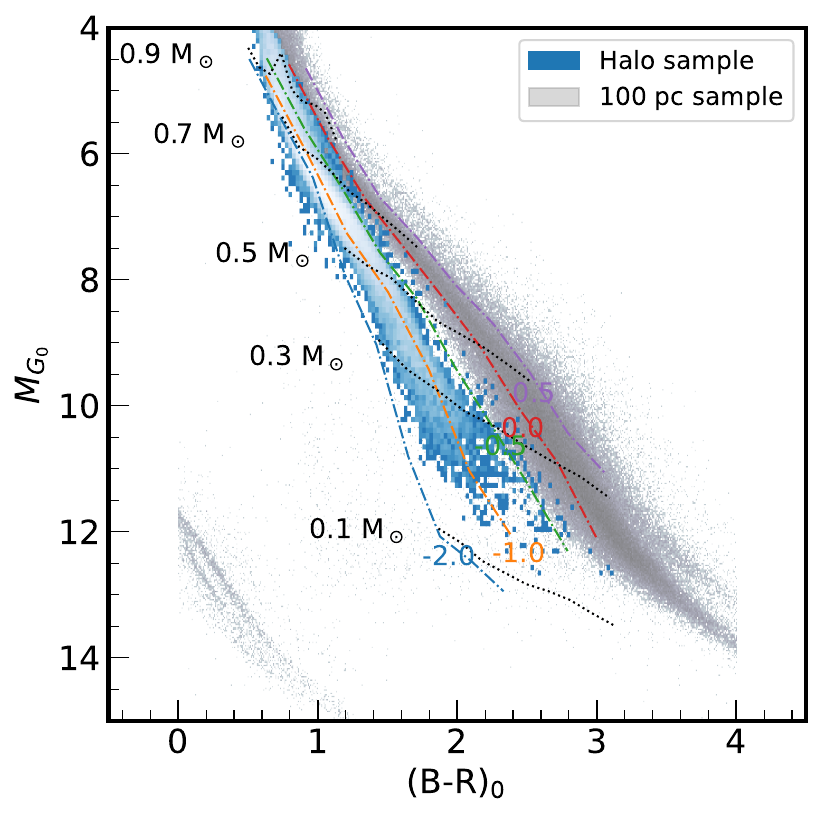}
    \caption{
    Left: Metallicity distribution of the kinematically pre-selected sample (halo-like orbits, $\eta>10$) and the final metal-poor sample. 
    The orange histogram shows the [Fe/H] distribution for the kinematically selected parent sample within 1 kpc, while the blue histogram shows the distribution for our final metal-poor sample, selected using both kinematic ($\eta$) and chemical ($\text{[Fe/H]} < -1.1$) criteria. The vertical dashed line indicates our metallicity cut.
    Right: The CMD for our final metal-poor sample with $\eta>10$ and [Fe/H]<-1.1.
    }
    \label{fig:halo_sample}
\end{figure*}

We take two steps to construct a high-purity sample of metal-poor stars, with [Fe/H]$<-1$: first, a we do a kinematic pre-selection to identify stars on halo-like orbits (which are predominantly metal-poor), followed by a spectroscopic metallicity cut to remove any remaining metal-rich contaminants. This section describes both steps in detail.
Further details of the halo sample construction and numbers are provided in Appendix~\ref{app:halo_selection}.

Stars on halo-like orbits are identified using a probabilistic selection based on orbital (tangential) kinematics combined with metallicity from Gaia spectrophotometric XP spectra \citep{DeAngeli2022, Montegriffo2023}.
The four-dimensional distribution formed by sky position and tangential velocity, $(\alpha,\delta,v_\alpha,v_\delta)$, is modeled conditional on spectrophotometric metallicity [Fe/H]. Here the tangential velocities are defined as $v_\alpha = \mu_\alpha/\varpi$ and $v_\delta = \mu_\delta/\varpi$, with $\mu_\alpha,\mu_\delta$ the proper motions and $\varpi$ the parallax. 
The conditional model $p(\alpha,\delta,v_\alpha,v_\delta \mid \text{[Fe/H]})$ separates disk-like and halo-like orbital behavior as a function of metallicity.
Although Gaia provides radial velocities for bright stars, only $\sim$1/3 of sources with $G<14$ have reliable RV measurements, and the fraction drops rapidly at fainter magnitudes, leading to a strongly magnitude-dependent selection if RVs were required.

The initial sample is constructed by querying the Gaia DR3 archive with the following cuts to ensure reliable astrometric measurements. Requirements include parallax $> 1$ mas (restricting the sample to stars within approximately 1 kpc), parallax-to-error ratio $> 10$ (\texttt{parallax\_over\_error} $> 10$), and absolute G-magnitude brighter than 17.5. The 1 kpc distance limit is adopted for two key reasons: (1) at this distance, main-sequence dwarf stars with masses down to $\sim 0.5~{\rm M}_{\odot}$ (corresponding to $M_G \sim 8$ mag) remain detectable within our magnitude limit, ensuring adequate coverage of the low-mass regime; and (2) the parallax signal-to-noise ratio is sufficiently high to provide reliable distance estimates for our spatial density modeling and mass estimate. Additionally, only sources with available XP continuous spectra are selected, yielding 27,978,851 stars in the initial sample.

The corresponding ADQL query is:
\begin{lstlisting}
SELECT * FROM gaiadr3.gaia_source 
WHERE has_xp_continuous = ``true''
  AND parallax > 1
  AND parallax_over_error > 10
  AND phot_g_mean_mag < 17.5
  AND phot_g_mean_mag + 5*log10(parallax/100)  > 4
\end{lstlisting}

Starting with the 27,978,851 stars from the initial query, a metal-poor stellar sample is constructed through a multi-stage selection process.  The de-reddening process follows the method of \cite{Li2025A}.
The \texttt{dustmaps} package \citep{2018JOSS....3..695M} is utilized to query a 3D dust map and derive the extinction parameter $E$.\footnote{The dust map provides $E$ in units of mag per parsec, which can be converted to extinction at any wavelength by applying the appropriate extinction curve.}  The average extinction curve from \cite{zhang2023}, combined with the transmission curves of the Gaia passbands, is adopted to compute the extinction corrections in B, G and R bands.
Extinction-corrected absolute magnitudes ($M_{G0}$) and colors $(B-R)_0$ are then computed.
To remove most binary systems and pre-main-sequence stars, an isochrone cut is applied by interpolating a [M/H] = $0$ isochrone from PARSEC \citep{Bressan2012} and selecting only stars below this metallicity threshold with $M_{G0} > 4$ mag, reducing the sample to 20,462,709 stars.

We quantified and cleaned problematic BP/RP photometry using the corrected excess factor $C^\ast$ from \cite{Riello2021}. For each star we computed the color B-R and
evaluated the cubic polynomial $f({\rm colour})$ with the Riello et al.\ coefficients $A_0$--$A_3$, and defined
\[
C^\ast = C - f({\rm B-R}),
\]
where $C$ is the catalog value \texttt{phot\_bp\_rp\_excess\_factor}.
We then estimated the expected scatter of $C^\ast$ as a function of $G$ using
\[
\sigma_{C^\ast}(G) = 0.0059898 + 8.817481\times10^{-12}\,G^{7.618399},
\]
and rejected stars with $|C^\ast|$ larger than a few times $\sigma_{C^\ast}(G)$, i.e.\ clear outliers with unreliable BP/RP fluxes. This cut removed about $7\%$ of the isochrone-selected sample, leaving $18{,}996{,}706$ stars out of the original $20{,}462{,}709$.
We note that the BP/RP excess-factor cut is not strictly CMD-neutral, but primarily removes sources with unreliable $G$ versus BP/RP photometry (e.g. very red, crowded, or highly extincted objects such as inner-bulge giants; \citealt{Riello2021}); this does not affect the CMD region or mass–metallicity regime analysed here.

We further refined the sample by requiring halo-like tangential kinematics, quantified by an orbital discriminator $\eta$ (Section~\ref{subsec:kinematic_model}; Appendix~\ref{app:gmm}), and applied an empirical threshold $\eta>10$ to suppress metal-rich disk contamination. 
We then derived metallicities from Gaia XP spectra using J-CaPS (Section~\ref{subsec:metallicity}; Appendix~\ref{app:jcaps_feh}) and imposed basic astrometric quality cuts using the Gaia EDR3 spurious-source catalogue \citep{Rybizki2022} (\texttt{fidelity\_v2}>0.5), yielding 253{,}247 candidates. 
Finally, applying [Fe/H]$<-1.1$ and $G<17$ produced our final sample of 53{,}275 stars spanning $\sim0.2$--$\sim0.9\,{\rm M}_\odot$ (Table~\ref{tab:sample_selection}). 
We adopt [Fe/H]$<-1.1$ as a conservative halo cut around the canonical [Fe/H]$\simeq-1$ division between rotationally supported and kinematically hot components \citep[e.g.][]{Chandra2024}, and $G<17$ as a conservative limit relative to the nominal XP depth ($G\simeq17.6$) to reduce magnitude-/colour-dependent XP systematics.

\subsection{The Probabilistic Kinematic Model}\label{subsec:kinematic_model}

The conditional 4D density, $p(\alpha,\delta,v_\alpha,v_\delta \mid \text{[Fe/H]})$, is modeled using a Gaussian Mixture Model (GMM). A GMM provides a flexible, semi-parametric representation of a probability density as a weighted sum of multivariate Gaussians:
\begin{equation}
p(\mathbf{4D}) = \sum_{k=1}^K \pi_k \,\mathcal{N}(\mathbf{4D}\mid\boldsymbol\mu_k,\boldsymbol\Sigma_k),
\end{equation}
where $\mathbf{4D}=(\alpha,\delta,v_\alpha,v_\delta)$ is the kinematic data vector. Full covariance matrices, $\boldsymbol\Sigma_k$, are adopted, enabling the model to capture complex correlations between the phase-space dimensions.  The detailed architecture and training of this model are presented in Appendix~\ref{app:gmm}.

The model is trained on a high-quality benchmark sample of red-giant-branch (RGB) stars selected from the Gaia XP catalog of \citep{Li2024, jiadong_li_2024_10469859}.
A key assumption of our approach is that the relationship between orbital kinematics and metallicity is independent of stellar evolutionary stage and stellar mass. 
This means that a model trained on RGB stars can be applied to identify likely metal-poor main-sequence stars, including the low-mass stars that are the focus of this work.
This assumption is mitigated by the fact that the kinematic model is explicitly conditioned on metallicity and empirically captures the disk–halo separation at each [Fe/H], rather than adopting a fixed metallicity boundary. 
At the low metallicities relevant here, contamination from genuinely young main-sequence stars is expected to be negligible.
These RGB stars are ideal tracers for calibrating this kinematic-metallicity relation: they are intrinsically bright, enabling all-sky coverage; they span a wide metallicity range from $[{\rm Fe/H}] \sim -2.5$ to $+0.5$ \citep{Li2024}, encompassing all major Galactic stellar populations from the metal-poor halo to the metal-rich disk \citep{Chandra2024}.
The specific quality cuts used to define this training set are detailed in Appendix~\ref{app:gmm}.

These conditional sample number densities, $n(\mathbf{4D}$) now allow us to construct an effective orbit-based metallicity-discriminator, $\eta$:
\begin{equation}\label{eq:eta}
\eta \equiv \log_{10}\frac{n(\mathbf{4D}\mid \text{[Fe/H]}<-1)}{n(\mathbf{4D}\mid \text{[Fe/H]}>-1)}.
\end{equation}
In essence, $10^\eta$ tells us how much more likely a star at any $\mathbf{4D}$ position is to have $\text{[Fe/H]}<-1$ than $\text{[Fe/H]}>-1$. 
Consequently, higher $\eta$ values signify a stronger kinematic indication of low metallicity. 

We perform an independent validation of the discriminator $\eta$ by examining its behavior on the color-magnitude diagram (CMD) for main-sequence stars, as shown in Figure~\ref{fig:cmd_eta_feh}.
In Figure~\ref{fig:cmd_eta_feh}, stars are color-coded by their $\eta$ value. Stars with higher $\eta$ values systematically occupy the bluer side of the main sequence.
This position corresponds to the location of older, more metal-poor populations according to stellar isochrones.
This demonstrates that the purely kinematic discriminator identifies stellar populations with distinct chemical properties, providing independent support for our kinematic pre-selection approach.

\subsection{Metallicity of Gaia XP}\label{subsec:metallicity}

To derive metallicities ([Fe/H]) from low-resolution Gaia XP spectra, the \textsl{J-CaPS} (JAX-based stellar parameter fitting code) framework is employed (Li et al., in prep).  This method uses neural networks to forward model the XP spectra given stellar atmospheric parameters ($T_{\text{eff}}$, $M_G$, and [Fe/H]). The model is trained on 76,632 stars with both Gaia XP spectra and ASPCAP stellar parameters from the SDSS-V/APOGEE IPL-4 catalog \citep{SDSS2025, Szabolcs2025, Kollmeier2025}.  This learned relationship is then inverted using \texttt{BFGS} optimization to estimate parameters for any given XP spectrum.
Cross-validation tests indicate a typical metallicity uncertainty of $\sim$0.25~dex in the [Fe/H]~$<$~$-$1 regime.
More details can be found in Appendix~\ref{app:jcaps_feh}.

The XP-derived [Fe/H] is validated in the CMD as shown in Figure~\ref{fig:cmd_eta_feh}, where the metallicity distribution follows the gradient of PARSEC evolutionary tracks: metal-rich stars ([Fe/H] $\gtrsim -0.5$) in the 100 pc (gray background) populate the red main sequence above the [M/H] $= 0.5$ isochrone, while metal-poor stars ([Fe/H] $\lesssim -1$) occupy the blue side.

\section{Modeling the Mass Function}\label{section:methods}

To infer the intrinsic stellar mass function from our observed sample, a statistical framework similar to \citet{Rix2021} is adopted. The core of this method is to model the number of stars detected in our catalog as a Poisson point process. This approach allows for accounting of observational biases, most notably the survey selection function and spatial incompleteness, by connecting an intrinsic physical distribution of the stellar population to the expected number of observed stars.

Throughout this work, the measurement is referred to as the \emph{stellar mass function} (MF) rather than the stellar \emph{initial mass function} (IMF). 
The IMF, in its strict definition, describes the distribution from which \emph{all} stars---both single and binary components---are drawn at birth \citep{kroupa1993, Kroupa2013, Wang2025}. 
However, our observational constraints and modeling approach are based on fitting single-star models to Gaia XP spectrophotometric data. This methodology cannot directly account for unresolved binary systems, which may have different mass distributions. Therefore, what is measured is more accurately described as the present-day mass function of the observable population, rather than the intrinsic IMF that includes both single and binary star formation channels.

\subsection{Forward Model of the Stellar Number Counts}

The goal is to determine the intrinsic mass function (MF), $\Phi_0(M)$, defined as the number of stars per unit mass per unit volume in kpc$^3$. The mass function $\Phi_0(M)$ is modeled non-parametrically as a piecewise-constant function, assuming a constant value $\Phi_k$ within each discrete mass bin $k$. The number of stars observed in our sample, $n_k$, is modeled as a draw from a Poisson distribution whose expected rate $\lambda_k$ is given by the model:
\begin{equation}
n_k \sim \mathrm{Poisson}(\lambda_k) \quad \text{where} \quad \lambda_k \approx \Phi_k \cdot \tilde{V}_{\mathrm{eff},k} \cdot \Delta M_k
\end{equation}
Here, $\tilde{V}_{\mathrm{eff},k}$ is the pre-computed numerical effective volume for that bin, $\Delta M_K$ is the width of the mass bin.
In a logarithmic framework, the effective volume term serves as a known offset that corrects the raw counts for selection biases:
\begin{equation}
\log \lambda_k = \log \Phi_k + \log(\tilde{V}_{\mathrm{eff},k} \cdot \Delta M_k).
\end{equation}

To connect this intrinsic function to observations, the spatial distribution of these stars within the Galaxy must also be modeled. The total number of stars at a given location is modeled as the product of the intrinsic MF and a spatial density profile, $n(\boldsymbol{x})$. The spatial density profile $n(x)$ is described by a standard cylindrical model:
\begin{equation}\label{eq:spatial_density}
n(R,Z) = n_0 \exp\left(-\frac{R-R_\odot}{h_R} - \frac{|Z|}{h_z}\right)
\end{equation}
where $(R, Z)$ are Galactocentric cylindrical coordinates, $R_\odot$ is the solar radius, and $h_R$ and $h_z$ are the radial and vertical scale heights, respectively.
We adopt $h_R$=1.5, and $h_z=1.43$ kpc from \cite{Xiang2025}, as typical density profiles for the metal-poor stellar population.

\subsection{The Selection Function and Predicted Observations}

Not every star within our survey volume ends up in our final catalog. 
The process of observation and sample creation is captured by a selection function, $S_{\mathcal{C}}(q)$, which gives the probability that a star with observable properties $q$ is included in our sample $\mathcal{C}$.
For our work, the primary observable property is the apparent magnitude, $G$.

The total selection function can be viewed as the product of the parent catalog's completeness and any additional user-defined sample cuts. 
Since Gaia XP spectra are limited to sources brighter than $G \sim 17.5$, we apply a hard cut at $G = 17.0$ to ensure uniform completeness across our sample.
For this analysis, we therefore adopt a simple step function for our selection function based on the apparent $G$-band magnitude:
\begin{equation}
S_{\mathcal{C}}(G) = 
\begin{cases} 
1 & \text{if } G \le 17.0 \\
0 & \text{if } G > 17.0 
\end{cases}
\end{equation}
This function implies that stars brighter than our limit are included, while stars fainter than it are excluded.
Here we ignore color selection, as Gaia shows no color bias in this regime \citep{Cantat-Gaudin2023}.
We note that astrometric and photometric quality cuts may introduce a mild, implicit color-dependent selection that is difficult to model, but primarily remove sources with unreliable measurements and do not affect the CMD region or mass–metallicity regime analysed here.

To simplify this prediction, we marginalize over the spatial dimensions, which are nuisance parameters for the purpose of inferring the mass function.
This leads to the concept of an effective volume, $V_{\mathrm{eff}}(M)$:
\begin{equation}
V_{\mathrm{eff}}(M) \equiv \int n(x) \, S_{\mathcal{C}}(q(M, x)) \, \mathrm{d}^3x
\end{equation}
This term is not a simple geometric sphere; rather, it represents the total volume, weighted by the stellar density, from which a star of mass $M$ would be bright enough to pass our magnitude cut and be included in our sample. 
Because intrinsically brighter (more massive) stars can be seen to greater distances before their apparent magnitude drops below $G=17$, the effective volume is a function of mass or absolute magnitude, as shown in Figure~\ref{fig:veff}. 
It encapsulates all spatial and observational biases into a single, mass-dependent term.
With this definition, the predicted count in each mass bin simplifies to its final form:
\begin{equation}
\lambda_k = \int_{M_k}^{M_{k+1}} \Phi_0(M) \, V_{\mathrm{eff}}(M) \, \mathrm{d}M.
\end{equation}
In practice, we used NUTS (No-U-Turn Sampler) \citep{Hoffman2011} via \texttt{NumPyro} \citep{Phan2019} to infer the $\log \Phi_k$ :
\begin{equation}
p(\{\log \Phi_k\} \mid \{n_k\}, \{\tilde{V}_{\rm eff, k}\}),
\end{equation}
Exponentiating the samples gives the posterior distribution over $\Phi_k(M)$.
In practice, both the Poisson likelihood for the observed number counts and the effective-volume calculation are carried out in joint \((M,\,[\mathrm{Fe/H}])\) bins, such that metallicity enters the selection function explicitly and \(V_{\rm eff}\) is evaluated independently for each mass--metallicity bin.

\subsection{Numerical Estimation of the Effective Volume}

The analytical integral for the effective volume, $V_{\mathrm{eff}}(M)$, is generally intractable.
This is because the selection function $S_{\mathcal{C}}(q)$ depends on complex survey limits.
Therefore, we numerically estimate this quantity using a Monte Carlo approach, denoting the estimate as $\tilde{V}_{\mathrm{eff}}(M)$.
We note that in practice, we compute $\tilde{V}_{\rm eff}$ independently for each (mass, [Fe/H]) bin, enabling us to infer the metallicity-dependent mass function.
For each mass $M$ on a predefined grid, we simulate a large population of stars, distributing them throughout the Galaxy according to our spatial density model $n(x)$ in Eq.~\ref{eq:spatial_density}. 
For each simulated star, we calculate its observable properties $q$ and apply our selection function to determine if it would be included in the final catalog ($S_{\mathcal{C}}(q) = 1$). 
The estimated effective volume is then the total volume of the simulation, weighted by the fraction of simulated stars that were successfully ``detected.'' 

\begin{figure}
\centering
\includegraphics[width=0.9\linewidth]{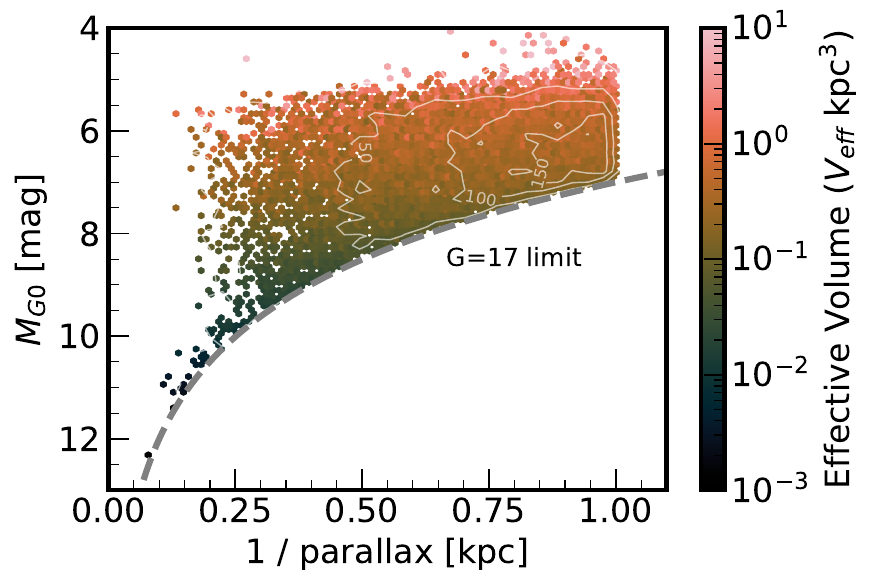}
\caption{The estimated effective volume $\tilde{V}_{\mathrm{eff}}$ as a function of absolute magnitude $M_{G0}$. The plot shows how the volume accessible to the survey is much larger for brighter (lower $M_{G0}$), more massive stars than for fainter, less massive ones.}
\label{fig:veff}
\end{figure}

The result of this computation is shown in Figure~\ref{fig:veff}, which illustrates how the survey's accessible volume changes as a function of stellar absolute magnitude (a proxy for mass). 
The uncertainties of $\tilde{V}_{\mathrm{eff}}(M)$ from
the Monte Carlo sampling are propagated into the likelihood via $\tilde{V}_{\mathrm{eff},k}$.

\section{Results}\label{sec:results}

\begin{figure*}[p]
    \centering
    \includegraphics[width=0.8\linewidth]{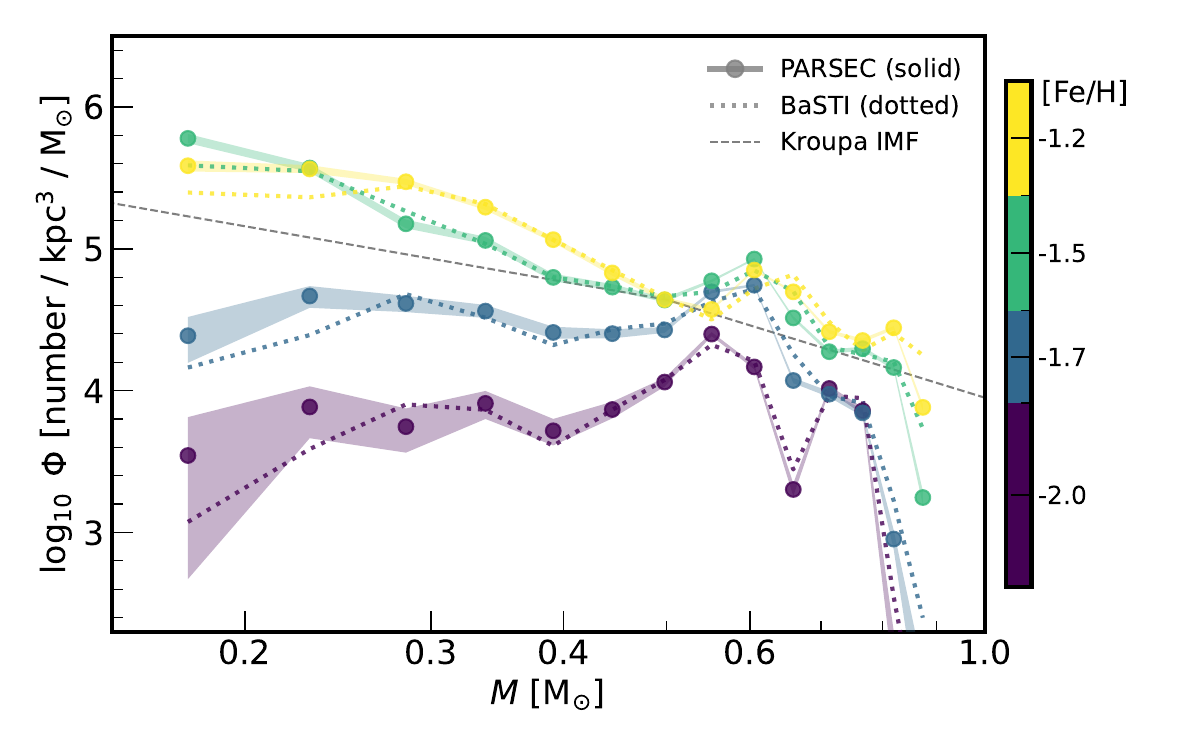}
    \caption{
    Stellar mass functions for the metal-poor sample.
    The logarithmic mass density, $\log_{10}\Phi$ [number\,kpc$^{-3}$\,M$_\odot^{-1}$], is shown as a function of stellar mass $M$ for different metallicity bins (color-coded by [Fe/H], see color bar).
    Stellar masses are derived from PARSEC models (solid lines) and BaSTI models (dotted lines).
    Symbols show the binned densities with shaded bands indicating $1\sigma$ uncertainties.
    The dashed gray line shows the canonical Kroupa IMF, scaled for visual comparison.
    }
    \label{fig:smf}
    \vspace{1cm}

    \includegraphics[width=0.99\linewidth]{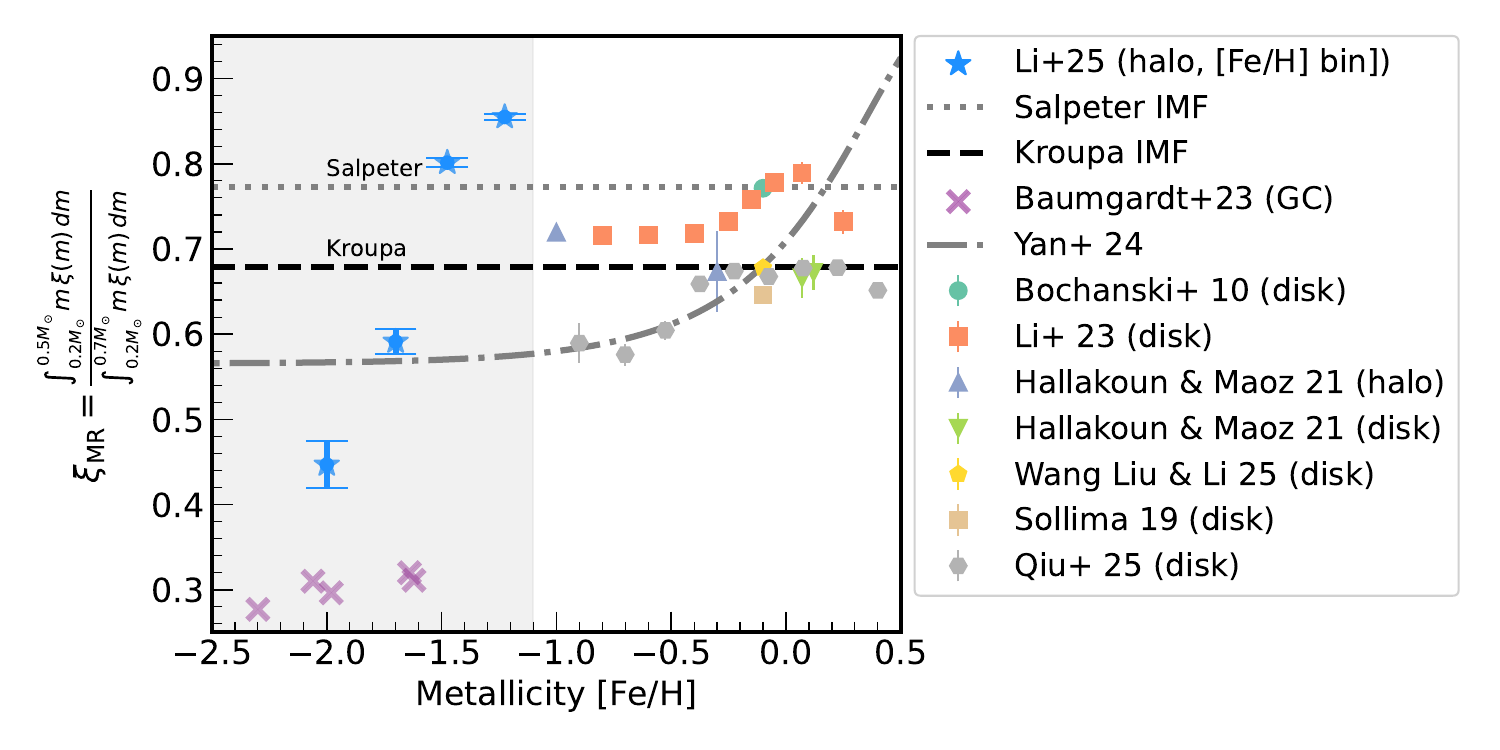}
    \caption{The mass ratio, $\xi_{\mathrm{MR}}$, as a function of [Fe/H].
    The ratio $\xi_{\mathrm{MR}}$ is defined as the stellar mass fraction between $0.2\,{\rm M}_\odot$ and $0.5\,{\rm M}_\odot$ relative to the mass between $0.2\,{\rm M}_\odot$ and $0.7\,{\rm M}_\odot$.
    Our results for metal-poor stars are shown as blue symbols for individual metallicity bins (stars).  These are compared against a compilation of literature data (other colored symbols). The horizontal black-dashed and grey-dotted lines represent the values for canonical Kroupa and Salpeter IMFs, respectively. The metallicity-dependent model from \cite{yan2024} is shown as a grey dash-dot curve. The shaded grey area shows the metal-poor regime with [Fe/H]$<-1.1$.}
    \label{fig:xi_MW}
\end{figure*}

\subsection{Stellar Mass Function}

We present our inferred stellar MF for the metal-poor population in Figure~\ref{fig:smf}.
The MF is derived as the posterior mean of the expected number from the Poisson model, $\Phi_k$, which represents the intrinsic number of metal-poor stars within each mass bin after volume correction.

Figure~\ref{fig:smf} shows the metal-poor mass function as a function of metallicity, revealing a clear metallicity-dependent trend.
We parameterize the MF below 0.55\,${\rm M}_\odot$ with a single power-law slope $\alpha_1$ (where $\mathrm{d}N/\mathrm{d}M \propto M^{\alpha_1}$). The four metallicity bins span [Fe/H]~=~-2.0, -1.7, -1.5, and -1.2, with measured slopes $\alpha_1 = 0.0 \pm 0.8$, $0.0 \pm 0.5$, $-2.7 \pm 0.2$, and $-1.8 \pm 0.4$.
The metallicity bin edges are set at [Fe/H] = $-2.2,\,-1.8,\,-1.6,\,-1.4,$ and $-1.1$, corresponding to bin widths of 0.4, 0.2, 0.2, and 0.3 dex. 
This choice reflects a trade-off between metallicity resolution and Poisson noise.

The most metal-poor bin ([Fe/H]~$\sim$~-2.0) has a near-flat slope ($\alpha_1 \sim 0$), indicating a flat MF with a deficit of low-mass stars below $0.5\,{\rm M}_\odot$. The [Fe/H]~$\sim$~-1.7 bin shows a similar flat trend ($\alpha_1 \sim 0$). In contrast, the [Fe/H]~$\sim$~-1.5 bin exhibits a steep negative slope ($\alpha_1 = -2.7 \pm 0.2$), steeper than the Kroupa IMF ($\alpha_{\rm Kroupa} = -1.3$ below 0.5\,${\rm M}_\odot$), indicating a bottom-heavy MF. The [Fe/H]~$\sim$~-1.2 bin has $\alpha_1 = -1.8 \pm 0.4$, also indicating a bottom-heavy MF but less extreme than the [Fe/H]~$\sim$~-1.5 bin. The transition between the flat ([Fe/H]~$\lesssim$~-1.7) and steep ([Fe/H]~$\gtrsim$~-1.5) regimes marks a change in the low-mass IMF.

Between 0.55 and 0.8\,${\rm M}_\odot$, we measure power-law slopes $\alpha_2 = -0.4 \pm 2.2$, $-2.8 \pm 2.0$, $-1.3 \pm 1.6$, and $-2.5 \pm 2.9$ across the same metallicity range. The large uncertainties in $\alpha_2$ reflect the limited dynamic range in this mass interval, where a single power-law does not provide a good fit to the observed MF.
The detailed mass function values $\Phi$ and their uncertainties for each metallicity bin are provided in Table~\ref{tab:mf_feh} in Appendix~\ref{app:mf}.

\subsection{Comparison with Literature}

To facilitate a meaningful comparison between measurements employing various formulations and modelings of the IMF, we adopt the parameter $\xi_\mathrm{MR}$ to represent the mass ratio of stars within specific mass ranges. Similar to \cite{MartinNavarro2019}, we define $\xi_\mathrm{MR}$ as the ratio of stellar mass in the range 0.2--0.5\,M$_\odot$ to that in the range 0.2--0.7\,M$_\odot$. 
Specifically, for a stellar IMF defined by $\xi(m)$, $\xi_\mathrm{MR}$ is expressed as:

\begin{equation}
    \xi_\mathrm{MR} \equiv \frac{\int_{0.2\,\mathrm{M}_\odot}^{0.5\,\mathrm{M}_\odot} m\,\xi(m)\,dm}{\int_{0.2\,\mathrm{M}_\odot}^{0.7\,\mathrm{M}_\odot} m\,\xi(m)\,dm}
\end{equation}
This parameter allows for a standardized comparison across different IMF formulations and observational datasets, revealing constraints into the low-mass end of the IMF and its potential variations with metallicity.

The canonical IMF (Kroupa IMF) yields a $\xi_\mathrm{MR}$ value of $0.67\pm0.05$, which serves as a benchmark for comparison.
Star-counting studies of the Milky Way disk by \citet{Sollima2019} and \cite{Wang2025}, which do not distinguish metallicity, fall within two standard deviations (2$\sigma$) of this value, suggesting consistency with the Canonical IMF for the integrated disk population.

In our previous study, \citet{Li2023} used LAMOST data to demonstrate a positive correlation between $\xi_\mathrm{MR}$ and metallicity for disk stars, indicating an increasing proportion of low-mass stars with rising metallicity.
Here we extend this analysis to the metal-poor regime ([Fe/H] $< -1$), where we find that this trend holds: metal-poor stars exhibit lower $\xi_\mathrm{MR}$ values as the metallicity decreases, indicating a bottom-light mass function.
As metallicity increases from [Fe/H] $\sim -2.0$ to $-1.1$, $\xi_\mathrm{MR}$ increases from $\sim$0.42 to $\sim$0.86, indicating a transition from bottom-light to bottom-heavy relative to the Kroupa value of 0.67. Our measurement at [Fe/H] $\sim -1.1$ shows agreement with the metal-poor IMF measurement by \citet{Hallakoun2021}, providing independent confirmation of an enhanced low-mass population at this metallicity.
Both the disk and metal-poor trends are consistent with expectations from the IGIMF model \citep{yan2024}, which adopts a metallicity-dependent low-mass IMF calibrated to reproduce a wide range of observational constraints, including star counts in metal-poor dwarf systems and integrated-light measurements of metal-rich galaxies. 
While this model captures the overall direction of the trend, the amplitude of the observed variation appears larger at the metallicity extremes.

Recent studies of the stellar IMF in globular clusters reveal a similar lack of low-mass stars in metal-poor environments \citep{Baumgardt2023, Dickson2023}.
We show in Figure~\ref{fig:xi_MW} the mass function of five globular clusters measured by \cite{Baumgardt2023}, selected to have undergone minimal dynamical evolution.
These clusters all have relaxation times larger than their ages and should therefore have experienced minimal mass loss due to escaping stars. The low level of mass segregation in these clusters also means that any mass loss they may have experienced is unlikely to bias their IMFs.
While these works found no clear trend between the IMF of clusters and metallicity within their sample, the qualitative agreement between the bottom-light IMFs of metal-poor GCs and our results for metal-poor field stars provides further, independent evidence that the IMF is bottom-light at low metallicities.

However, the offsets are evident among different studies, particularly for solar-abundance stars. The results of \citet{Bochanski2010} and \citet{Li2023} suggest a more bottom-heavy IMF (higher $\xi_\mathrm{MR}$) compared to \textit{Gaia}-only studies \citep{Sollima2019, Hallakoun2021} and \cite{Wang2025} at similar metallicities. 
These discrepancies underscore the importance of considering survey-specific selection functions, binary star corrections, and other potential systematic effects in IMF determinations.
The metal-poor population presents a particularly intriguing case.

Our findings are consistent with the IGIMF model trend for low-metallicity environments \citep{yan2024}. 
The model's success in capturing this trend is notable, given that it was calibrated using diverse data sources, including Galactic Chemical Evolution modeling and unresolved stellar systems.
In the metal-poor regime ([Fe/H]<-1),  the IGIMF model predicts only a weak variation, while the direction of the observed trend is consistent with a progressively bottom-light IMF toward lower metallicity.

\section{Discussion and Conclusion}

\subsection{Summary}

We have measured the stellar mass function of metal-poor stars ([Fe/H] $< -1.1$) within 1~kpc of the Sun from \textit{Gaia} DR3.
By combining probabilistic kinematic selection via Gaussian Mixture Models with XP-based spectrophotometric metallicities (calibrated to SDSS-V/ASPCAP) and explicit metallicity cuts, we constructed a high-purity metal-poor sample of $\sim$53,000 stars.
Using our forward-modeling approach that accounts for the \textit{Gaia} selection function, we find evidence for a metallicity-dependent mass function in the low-mass regime.

Our key finding is that the MF below 0.55\,${\rm M}_\odot$ transitions with metallicity: from a near-flat power-law slope ($\alpha_1 \sim 0$) at [Fe/H]~$\sim$~-2 to a steep slope ($\alpha_1 \sim -1.8$) at [Fe/H]~$\sim$~-1.
This corresponds to a change in the mass ratio $\xi_{\rm MR}$ from $\sim$0.42 to $\sim$0.86, indicating a transition from bottom-light to bottom-heavy relative to the Kroupa IMF.
This result demonstrates that the low-mass stellar mass function is not universal, but varies systematically with the metallicity of the stellar population.

\subsection{Caveats}
We note several features in the derived mass functions around
0.6--0.7\,${\rm M}_\odot$ that warrant discussion. The [Fe/H]~$\sim$~$-2$
bin shows a dip at $\sim$0.65\,${\rm M}_\odot$, while the [Fe/H]~$\sim$~$-1.5$
and [Fe/H]~$\sim$~$-1.2$ bins show corresponding peaks near 0.6\,${\rm M}_\odot$.
These features likely arise from two sources.

\paragraph{Metallicity uncertainties and bin-scattering.}

Metallicity estimation uncertainties in the XP spectra
($\sigma_{\rm [Fe/H]} \sim 0.25$\,dex; see Appendix~\ref{app:jcaps_feh})
can cause stars near metallicity-bin boundaries to scatter between bins.
In particular, the most metal-poor stars ([Fe/H] $< -1.5$) may be shifted
into higher-metallicity bins, producing excesses at 0.6–0.7\,${\rm M}_\odot$
and a corresponding deficit in the lowest-[Fe/H] bin.

\paragraph{Stellar-model uncertainties.}
Uncertainties in stellar evolution models at these masses and metallicities
may contribute to features around 0.6--0.7\,${\rm M}_\odot$.
For instance, \citet{Chen2014} replaced the traditional gray atmosphere boundary conditions with $T$--$\tau$ relations from PHOENIX BT-Settl model atmospheres, empirically recalibrated to match observed mass-radius relations. Their calibration used ultra-deep \emph{HST}/ACS observations of the globular clusters NGC~6397 and 47~Tuc, spanning two dex in metallicity, which allows PARSEC v1.2S models to reproduce the lower main sequence in metal-poor clusters better than other widely used grids such as Dartmouth and PHOENIX models. 
However, \citet{Chen2014} noted that systematic discrepancies remain in the optical colors, highlighting ongoing uncertainties in bolometric corrections for broad-band filters such as Gaia passbands. These model uncertainties may contribute to the features we observe around 0.6--0.7\,${\rm M}_\odot$, particularly in the mass-to-color conversions at the lowest metallicities. 
To assess the robustness of our results to model choice, we repeated our analysis using BaSTI models and found consistent results for the low-mass regime below 0.5\,${\rm M}_\odot$ (see Figure~\ref{fig:smf}).

\paragraph{Potential metallicity systematics at [Fe/H] < $-1.5$.}
A further concern is a possible mass-dependent metallicity offset in the
J-CaPS XP-based [Fe/H] estimates. 
The training set contains only two stars with [Fe/H] $< -1.5$ and $M < 0.5\,M_\odot$, placing this regime at the edge of the label distribution. 
However, the relevant atmospheric parameter space is better sampled than this raw number suggests: stars between 0.2--0.5\,${\rm M}_\odot$ all have $T_{\rm eff} \approx 4000$\,K with a dispersion of $\sim$180\,K, giving $\sim$25 training stars in the effective temperature range that determines the spectral shape. Because the J-CaPS forward model maps $(T_{\rm eff}, M_G, \mathrm{[Fe/H]})$ to the XP spectrum, the only extrapolation occurs in $M_G$, which we verify scales nearly linearly with the XP flux. 
Thus a 0.3\,$M_\odot$ and a 0.6\,$M_\odot$ star with the same $T_{\rm eff}$ and [Fe/H] exhibit almost identical relative XP spectra, with the differences captured by a linear $M_G$ scaling.

Furthermore, the CMD color–coded by XP metallicity (Fig~\ref{fig:cmd_eta_feh}) shows no obvious evidence for a mass-dependent [Fe/H] bias: stars follow the expected metallicity sequence across the main sequence, consistent with PARSEC isochrones. 
Although extending the calibration set at low metallicity would be valuable in future works.

\subsection{Effect of Binarity}

Our analysis is based on fitting single-star models to the Gaia XP spectrophotometric data and does not explicitly account for unresolved binary systems.
However, several lines of evidence suggest that binarity does not significantly affect our primary conclusions about the metallicity-dependent mass function.

First, our J-CaPS fitting method uses absolute magnitude as a constraint when fitting XP spectra, which means that binary stars with large mass ratios (where the secondary contributes significantly to the total light) should produce poor fits to single-star models.
To quantify the prevalence of binaries in our sample, we cross-matched our metal-poor stars with the binary catalog from \citet{Li2025B}, which compares single-star and binary-star model fits to XP spectra.
We find that $\sim$15\% of our sample stars show better fits to binary models than to single-star models, suggesting that the majority of our sample consists of either single stars or binaries with small mass ratios (where the secondary contributes minimally to the observed light).

Second, we assess the binary fraction through radial velocity variability.
We cross-matched our metal-poor sample with multi-epoch spectroscopic observations from SDSS-V/BOSS, identifying 5,096 common stars with multiple radial velocity measurements.
Of these, only 45 stars ($\sim$0.9\%) show radial velocity variations $\Delta$RV larger than three times the typical BOSS measurement uncertainty ($\sim$5~km\,s$^{-1}$), indicating a low fraction of spectroscopic binaries with detectable orbital motion on the BOSS observational timescales.
This low detection rate is consistent with expectations for a population dominated by either single stars or wide binaries with long orbital periods.

Third, we performed a Monte Carlo simulation to assess how undetected binaries might bias our mass function determination.
We generated a mock population of 10,000 stars following a Salpeter IMF (power-law slope $\alpha = 2.35$) with a 40\% binary fraction, representative of solar-neighborhood populations \citep{Duchene2013}.
Both single stars and individual components of binary systems were drawn from the IMF.
We then selected only objects that are either genuinely single or binary systems with mass ratios $q < 0.5$ (where the secondary is sufficiently faint to not significantly affect photometry) and treated all selected objects as single stars in a star-counting analysis.
The recovered best-fit power-law slope is $\alpha = 2.34 \pm 0.05$, indistinguishable from the input Salpeter slope at the $1\sigma$ level.
This demonstrates that even with a substantial binary fraction, star-counting methods are relatively robust to binary contamination when the sample is dominated by single stars and small mass-ratio systems.

Taken together, these three independent checks suggest that binarity does not dominate the systematic uncertainties in our derived mass functions.
The metallicity-dependent trend we observe---from bottom-light at [Fe/H]~$\sim$~-2 to bottom-heavy at [Fe/H]~$\sim$~-1---is robust to the level of binary contamination present in our sample.
However, a comprehensive treatment of binary populations will be important for future precision studies.
Our future work will extend the metal-poor mass function analysis by explicitly modeling binary star populations using the methods developed by \citet{Wang2025}.

\subsection{Broader Implications}

Our findings are broadly consistent with predictions from the metallicity-dependent IGIMF model \citep{2018A&A...620A..39J,yan2024} in low-metallicity environments.
These results challenge the assumption of IMF universality and have implications for several areas of astrophysics.
First, they provide a natural explanation for the "M Dwarf Problem"---the deficit of low-mass, metal-poor stars relative to simple closed-box chemical evolution models.
Second, they suggest that star formation in the early Universe and in metal-poor satellite galaxies preferentially produced higher-mass stars, affecting interpretations of early galaxy stellar masses and star formation rates.
Third, the bottom-light metal-poor IMF may help reconcile the properties of the Milky Way's stellar halo with hierarchical galaxy formation models.

Future work extending comprehensive modeling of binary populations and their evolution will be important for connecting the observed present-day mass function to the primordial IMF.

\section{Acknowledgments}
JL thanks David W. Hogg, Dan Maoz and Sebastien Lepine for helpful discussions.
JL and HWR acknowledge support from the European Research Council through ERC Advanced Grant No. 101054731.
YST is supported by the National Science Foundation under Grant AST-2406729.
J.G.F-T gratefully acknowledges the support provided by ANID Fondecyt Regular No. 1260371, ANID Fondecyt Postdoc No. 3230001 (Sponsoring researcher), the Joint Committee ESO-Government of Chile under the agreement 2023 ORP 062/2023 and the support of the Doctoral Program in Artificial Intelligence, DISC-UCN.
This work has made use of data from the European Space Agency (ESA) mission {\it Gaia} (\url{https://www.cosmos.esa.int/gaia}), processed by the {\it Gaia} Data Processing and Analysis Consortium (DPAC, \url{https://www.cosmos.esa.int/web/gaia/dpac/consortium}). Funding for the DPAC has been provided by national institutions, in particular the institutions participating in the {\it Gaia} Multilateral Agreement.

Funding for the Sloan Digital Sky Survey V has been provided by the Alfred P. Sloan Foundation, the Heising-Simons Foundation, the National Science Foundation, and the Participating Institutions. SDSS acknowledges support and resources from the Center for High-Performance Computing at the University of Utah. SDSS telescopes are located at Apache Point Observatory, funded by the Astrophysical Research Consortium and operated by New Mexico State University, and at Las Campanas Observatory, operated by the Carnegie Institution for Science. The SDSS web site is \url{www.sdss.org}.

SDSS is managed by the Astrophysical Research Consortium for the Participating Institutions of the SDSS Collaboration, including the Carnegie Institution for Science, Chilean National Time Allocation Committee (CNTAC) ratified researchers, Caltech, the Gotham Participation Group, Harvard University, Heidelberg University, The Flatiron Institute, The Johns Hopkins University, L'Ecole polytechnique f\'{e}d\'{e}rale de Lausanne (EPFL), Leibniz-Institut f\"{u}r Astrophysik Potsdam (AIP), Max-Planck-Institut f\"{u}r Astronomie (MPIA Heidelberg), Max-Planck-Institut f\"{u}r Extraterrestrische Physik (MPE), Nanjing University, National Astronomical Observatories of China (NAOC), New Mexico State University, The Ohio State University, Pennsylvania State University, Smithsonian Astrophysical Observatory, Space Telescope Science Institute (STScI), the Stellar Astrophysics Participation Group, Universidad Nacional Aut\'{o}noma de M\'{e}xico, University of Arizona, University of Colorado Boulder, University of Illinois at Urbana-Champaign, University of Toronto, University of Utah, University of Virginia, Yale University, and Yunnan University.

\vspace{5mm}
\facilities{\textsl{Gaia}}


\software{
{\tt\string PyTorch} \citep{NEURIPS2019_9015},
{\tt\string JAX} \citep{jax2018github},
{\tt\string Astropy} \citep{2018AJ....156..123A}, 
{\tt\string Scipy} \citep{2020SciPy-NMeth}
{\tt\string Zuko} \citep{rozet2022zuko}
}



\clearpage
\appendix
\onecolumngrid

\section{Metal-poor sample selection}\label{app:halo_selection}

The selection of metal-poor stars from the Gaia dataset involves several steps to ensure a clean and representative sample. The selection criteria are listed in Table~\ref{tab:sample_selection}.

\begin{deluxetable}{llr}
\tablewidth{0pt}
\tablecaption{Data Selection Criteria for Metal-Poor Star Sample \label{tab:sample_selection}}
\tablehead{
\colhead{Selection Step} & \colhead{Criterion} & \colhead{Sample Size}
}
\startdata
Initial sample & --- & 27,978,851 \\
Isochrone cut & Below [M/H]\,$=$\,0 isochrone; $M_{G0} > 4$; $0 < (B-R)_0 < 4$ & 20,462,709\\
Halo-like orbits & $\eta > 10$ & 253,247 \\
Metal-poor cut & [Fe/H]$_{\rm J-CaPS} < -1.1$ & 70,346 \\
Magnitude limit & $G < 17$ mag & 53,275 \\
\enddata
\end{deluxetable}

\section{Gaussian mixture model training and validation }\label{app:gmm}

\subsection{Training sample: RGB benchmark}

As a high-quality training set we select red-giant-branch (RGB) stars from the Gaia XP catalogue of \citet{jiadong_li_2024_10469859}. The criteria are listed in Table~\ref{tab:rgb_selection} and yield a sample for training the kinematical GMM model.

\begin{table}[h]
\centering
\begin{tabular}{ll}
\hline
Criterion & Value \\
\hline
Metallicity error & $e_{\rm [Fe/H]_{xp}} < 0.1$ \\
Effective temperature error & $e_{\rm T_{\rm eff,xp}} < 50\ \mathrm{K}$ \\
Surface gravity error & $e_{\log g_{\rm xp}} < 0.1$ \\
S/N of RP & $>50$ \\
RUWE & $<1.4$ \\
S/N of parallax & $>10$ \\
\hline
\end{tabular}
\caption{Selection criteria for RGB stars used as the benchmark training set.}
\label{tab:rgb_selection}
\end{table}

\subsection{Model Training}
We train a conditional Gaussian Mixture Model (GMM) using the \texttt{zuko} library \citep{rozet2022zuko} to model the four-dimensional feature distribution. The number of components $K$ is chosen using standard model-selection criteria and by visual inspection of the fit; we settled on $K=20$.
The parameters of these components (weights, means, and covariances) are generated by a hyper-network conditioned on metallicity. This network consists of two hidden layers, each with 32 units.

The model's parameters, $\boldsymbol{\theta}$, are optimized by minimizing a weighted negative log-likelihood loss function. For a batch of $N$ data points $\{\mathbf{x}_i, c_i, w_i\}$, where $\mathbf{x}_i$ is a data point, $c_i$ is its context, and $w_i$ is its sample weight, the loss $\mathcal{L}$ is defined as:
\begin{equation}
    \mathcal{L} = - \frac{G}{N} \sum_{i=1}^N w_i \log p(\mathbf{x}_i | c_i; \boldsymbol{\theta}),
\end{equation}
where $p(\mathbf{x}_i | c_i; \boldsymbol{\theta})$ is the likelihood given by the GMM. A constant gain factor $G$ is set to $10^6$ for numerical scaling during training.

Optimization is performed using the \texttt{Adam} optimizer with an initial learning rate of $10^{-3}$. We employ the \texttt{ReduceLROnPlateau} learning rate scheduler, which reduces the learning rate by a factor of 0.5 if the validation loss shows no improvement for 5 validation cycles (checked every 20 epochs). The model is trained for a maximum of 1000 epochs, and the training is stopped early if the learning rate drops below $10^{-6}$.

\begin{figure}[htbp]
    \centering
    \includegraphics[width=0.7\linewidth]{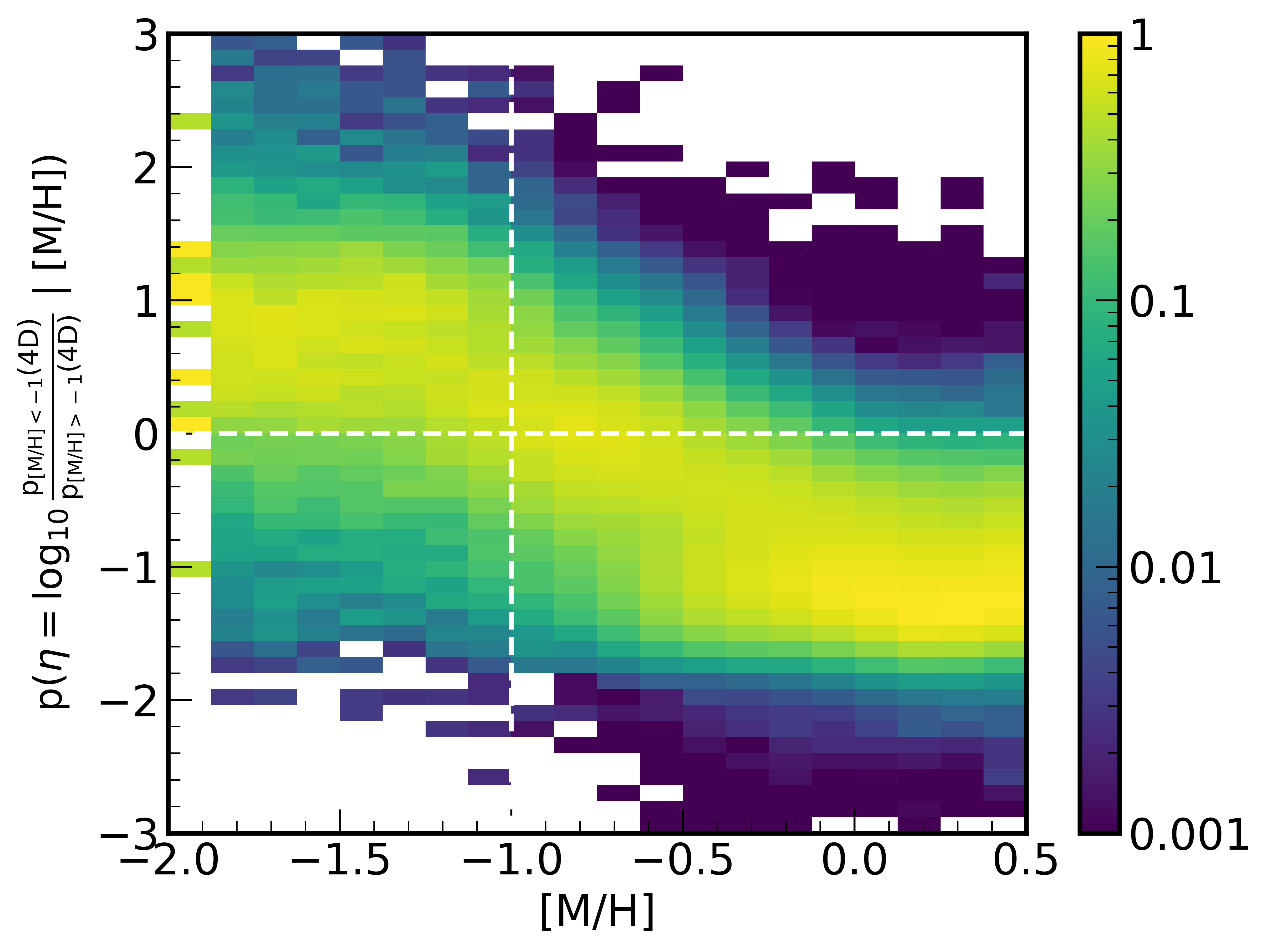}
    \caption{Validation of the orbital discriminator $\eta$.  The distribution of RGB benchmark stars in metallicity [Fe/H] vs. $\eta$ space.
    The color bar denotes the column-normalized number density.
    }
    \label{fig:eta}
\end{figure}

We validate the discriminator on the RGB stars with known metallicities from \cite{jiadong_li_2023_8002699}. 
Figure~\ref{fig:eta} shows the 2D distribution of stars in ([Fe/H], $\eta$) space, which demonstrates separation between metal-poor and metal-rich populations.

\section{Stellar Parameter Estimation Methods}

We require estimates of fundamental stellar parameters for each star in our sample, specifically mass and metallicity. 
This section details the data-driven methods we developed to derive these quantities. 
First, we present a neural network model trained on theoretical isochrones to estimate stellar masses (Subsection~\ref{app:mass_estimation}). 
Second, we describe our use of the \textsl{J-CaPS} framework, which performs spectral fitting on Gaia XP spectra to determine metallicities (Subsection~\ref{app:jcaps_feh}).

\subsection{Neural Network for Mass Estimation from Isochrones}\label{app:mass_estimation}

To efficiently derive stellar masses for large catalogs, we developed and trained a neural network (NN) to emulate the mapping from photometric observables to stellar mass. Our network is trained on a dense grid of PARSEC isochrones \citep{Bressan2012}, providing a robust theoretical foundation. 
We restricted the training data to stellar ages up to 10 Gyr and masses between 0.1 and 1.5 $M_{\odot}$. 
The choice of an NN architecture is motivated by the universal approximation theorem \citep{ Cybenko1989}, which ensures its capability to model the complex, non-linear relationship between a star's photometric properties and its mass.

The isochrone data points were partitioned into an 80-20 split for training and validation. The performance of the trained model is evaluated by the root-mean-square (RMS) error on the validation set. We achieve a high degree of precision, with an RMS scatter of 0.02 $M_{\odot}$ for mass and 0.1 dex for photometric metallicity ([Fe/H]), which confirms the high fidelity of our parameter estimates.

\subsection{\textsl{J-CaPS}: Metallicity from Data-Driven Spectral Fitting}\label{app:jcaps_feh}

To derive metallicities ([Fe/H]) directly from the low-resolution Gaia XP spectra, we employ the \textsl{J-CaPS} (JAX-based stellar parameter fitting code) framework.
This method uses deep neural networks and modern optimization techniques to extract stellar atmospheric parameters ($T_{\text{eff}}$, $M_G$, and [Fe/H]) from Gaia XP data.
The neural network architecture consists of four hidden layers with 512, 256, 128, and 64 neurons respectively, employing GELU (Gaussian Error Linear Unit) activation functions to model the nonlinear mapping from stellar parameters to observed XP spectra.

\begin{figure}[htbp]
    \centering
    \includegraphics[width=0.7\linewidth]{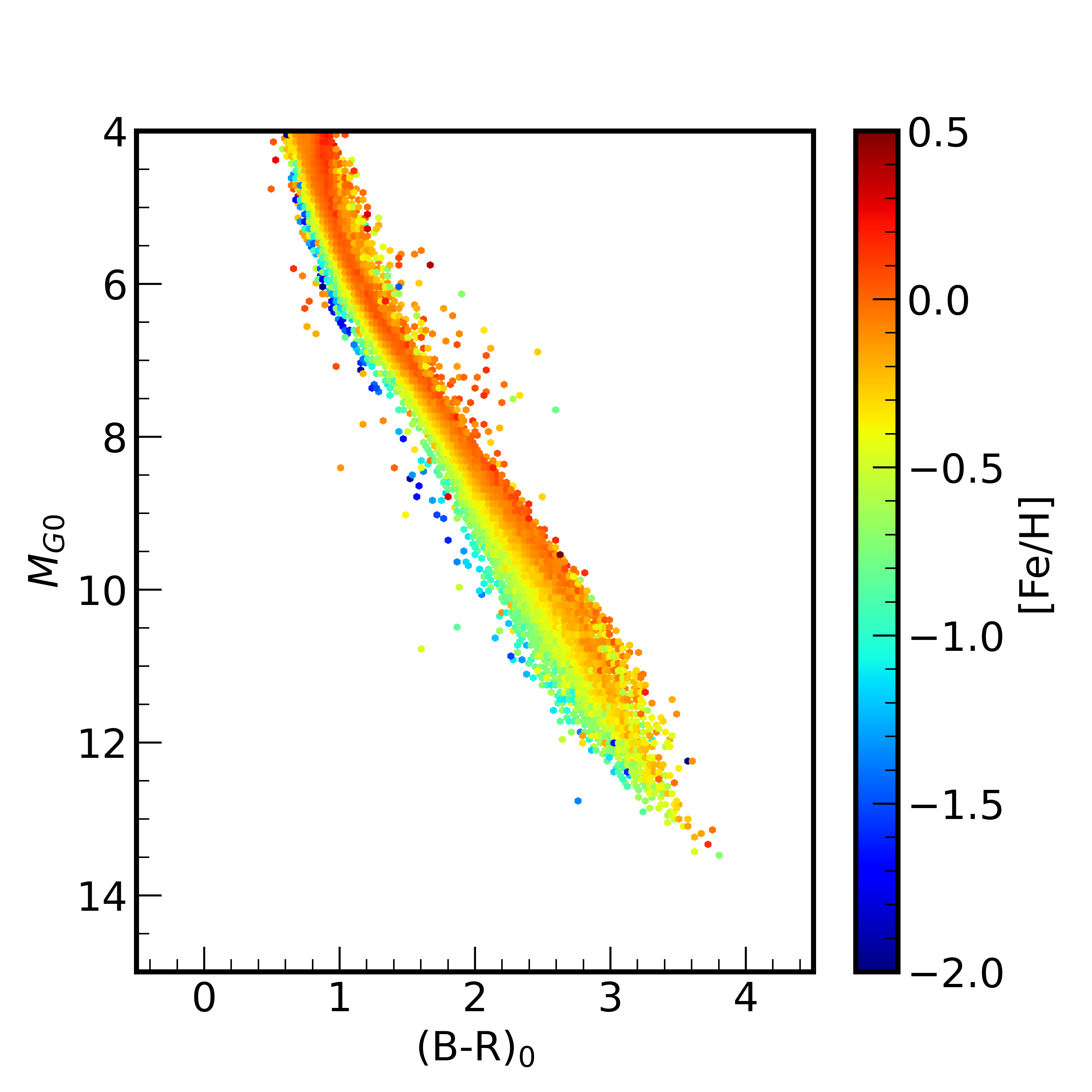}
    \caption{The CMD of Gaia sources cross-matched with SDSS-V/ASPCAP catalog. The color represents the [Fe/H] given by the ASPCAP catalog.}
    \label{fig:train}
\end{figure}

The core of J-CaPS is a data-driven spectral model, similar to \textsl{The Payne} \citep{Ting2019}.
We first train a forward model, implemented as a deep neural network, which learns the mapping from stellar parameters to observed XP spectra. 
This model is trained on a high-quality cross-matched sample of 76,632 stars with both Gaia XP spectra and reliable reference parameters from the SDSS-V/APOGEE survey \citep{Szabolcs2025}.
Table~\ref{tab:train_sample_selection} shows the details of the selection of the training data, and Figure~\ref{fig:train} shows the color-magnitude diagram of the training sample.
From the SDSS-V IPL-4 ASPCAP catalog we select stars with $\sigma_{\mathrm{[Fe/H]}} < 0.1$~dex, $\sigma_{T_{\mathrm{eff}}} < 100$~K, and set $\texttt{flag\_warn} = \texttt{flag\_bad} = 0$ to ensure that only stars with high-quality ASPCAP parameters are in the training set.
We also restricted $e_{v_{\mathrm{rad}}}$ and $\texttt{std\_v\_rad}$ values to be smaller than $1~\mathrm{km\,s^{-1}}$. Selecting stars with these values ensures that most variable stars are excluded from the sample, for which ASPCAP may provide incorrect parameters.

We use calibrated $T_{\mathrm{eff}}$, $\log g$ and [Fe/H] values instead of raw ones.
While the quality of uncalibrated $T_{\mathrm{eff}}$ is generally good in DR19 (median offset between $-60$ and $80$~K across seven different temperature scales; see \cite{Szabolcs2025} for details), it may have larger offsets for red dwarfs below $4500$~K.
This is because ASPCAP does not provide accurate surface gravities for these low-temperature main-sequence stars, and that may propagate into the temperature and abundances. On the red giant branch, the calibrated surface gravities also give much better agreement with the asteroseismic values.
We show the distribution of metal-poor 640 training stars across stellar masses of [Fe/H]$<-1$ in Table~\ref{tab:training_distribution}. We have 95 training samples with [Fe/H]$<-1.5$.

\begin{deluxetable}{lrl}
\tablewidth{0pt}
\tablecaption{Training sample selection criteria and resulting sizes \label{tab:train_sample_selection}}
\tablehead{
\colhead{Selection Step} & \colhead{N$_{\text{stars}}$} & \colhead{Criteria}
}
\startdata
Initial APOGEE & 437,056 & SDSS-V astraAllStarASPCAP-0.8 with $\log~g > 4$\\
Quality control & 179,620 & $\sigma_{[\text{Fe/H}]} < 0.1$ dex, $\sigma_{T_{\text{eff}}} < 100$ K, $\varpi/\sigma_\varpi > 5$,\\
 & & \texttt{flag\_warn}~$=$~\texttt{flag\_bad}~$=$~0, $e_{\text{v,rad}} < 1$ km/s, $0 < B - R < 4$ mag \\
XP cross-match & 159,942 & Matched with Gaia XP spectra \\
MSMS cross-match & 126,681 & Matched with MSMS catalog \cite{Li2025B} \\
Final single-star & 76,632 & $\chi^2_{\text{single}} - \chi^2_{\text{binary}} < 1$, $\chi^2(\text{single}) < 5$, $A_G < 0.1$ mag  \\
\enddata
\end{deluxetable}

\begin{deluxetable}{lrrrr}
\tablewidth{0pt}
\tablecaption{Training data distribution across mass and metallicity bins \label{tab:training_distribution}}
\tablehead{
\colhead{[Fe/H]} & \multicolumn{3}{c}{Mass [${\rm M}_\odot$]} & \colhead{Total} \\
\colhead{} & \colhead{$0.1$--$0.5$} & \colhead{$0.5$--$0.7$} & \colhead{$0.7$--$0.9$} & \colhead{}
}
\startdata
$-2.0$ to $-1.5$ & 2 & 7 & 86 & 95 \\
$-1.5$ to $-1.0$ & 24 & 48 & 473 & 545 \\
\hline
Total & 26 & 55 & 559 & 640 \\
\enddata
\end{deluxetable}

\begin{figure}[htbp]
    \centering
    \includegraphics[width=\linewidth]{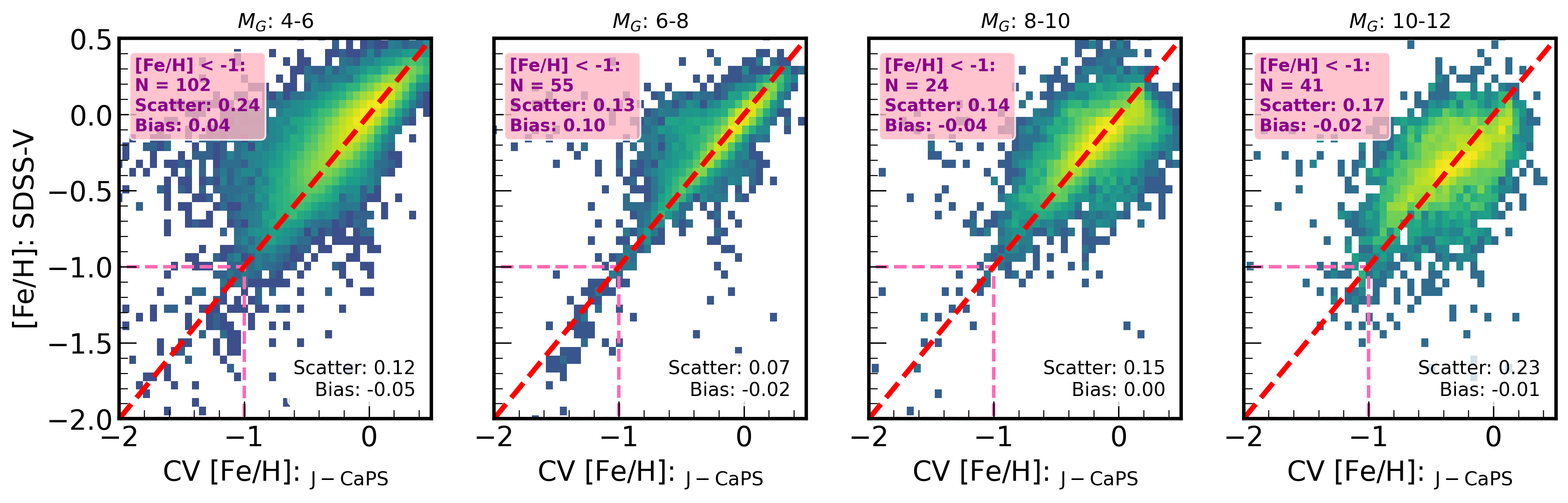}
    \caption{Validation of our 3-fold cross-validated (CV) metallicity estimates derived from XP (J-CaPS) against ASPCAP reference values from the SDSS-V.
    Each panel displays a 2D density histogram, comparing the J-CaPS $[Fe/H]$ estimate (x-axis) to the SDSS-V $[Fe/H]$ value (y-axis) for a different range of absolute Gaia magnitude ($M_G$), indicated at the top of the panel. The color scale represents the number density of stars. The red dashed line shows the ideal one-to-one correspondence.}
    \label{fig:cv_feh}
\end{figure}

\begin{figure}[htbp]
    \centering
    \includegraphics[width=1\linewidth]{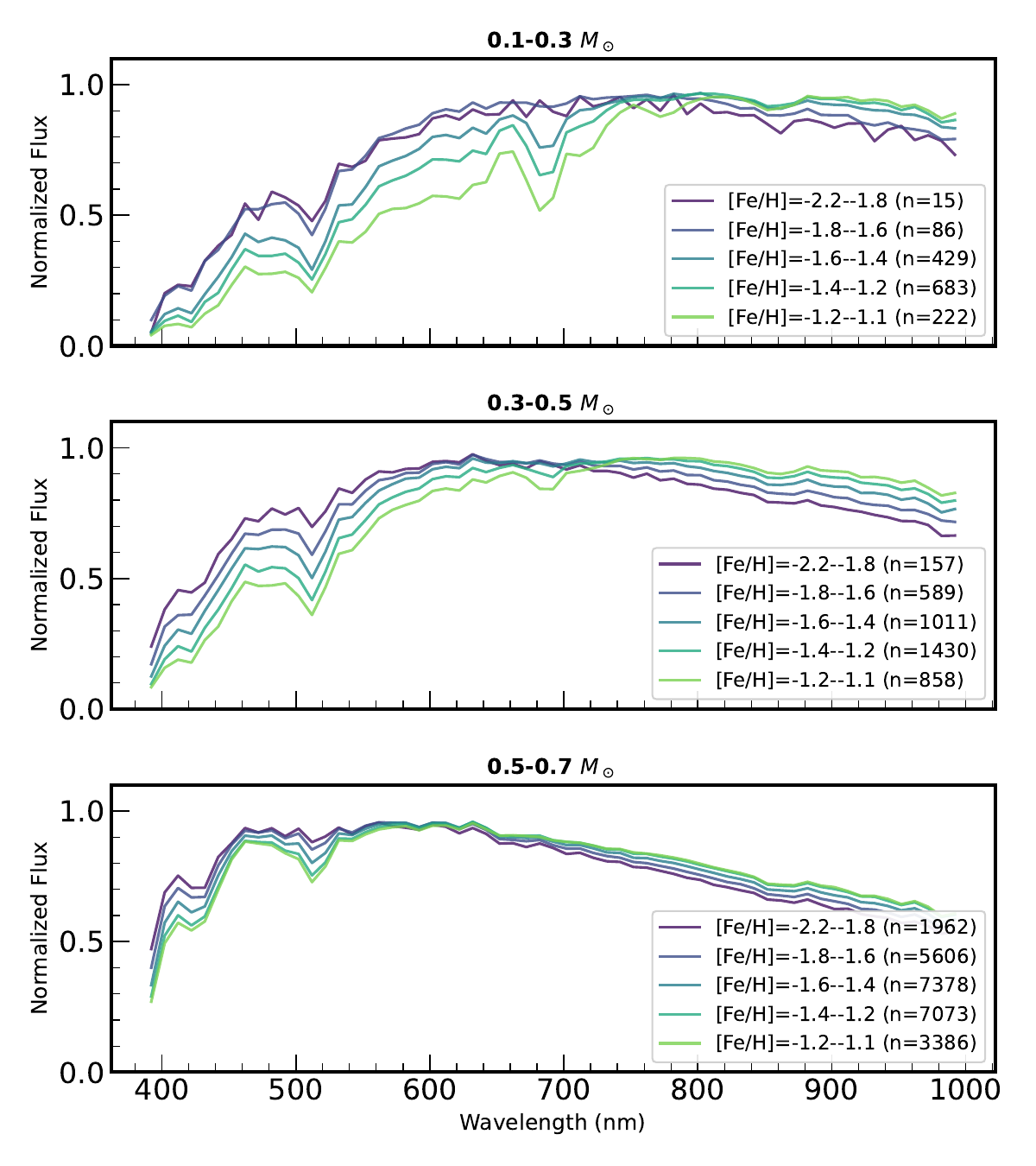}
    \caption{Median Gaia XP spectra for the metal-poor sample binned by stellar mass and metallicity. 
    The panels show three mass bins (0.1--0.3 $M_\odot$, 0.3--0.5 $M_\odot$, and 0.5--0.7 $M_\odot$) with spectra color-coded by metallicity [Fe/H]. Each spectrum represents the median normalized flux of stars in the corresponding mass-metallicity bin.}
    \label{fig:xp_median_spectra}
\end{figure}

Subsequently, this learned relationship is inverted using \texttt{BFGS} optimization algorithms to perform parameter estimation on any given observed XP spectra.
The parameter estimation itself proceeds in two stages to ensure stability and accuracy. 
First, a multi-layer perceptron provides an initial "best-guess" for the stellar parameters directly from the input XP spectrum. 
Second, these initial estimates are refined by the forward model in an optimization loop that minimizes the chi-squared ($\chi^2$) difference between the observed and predicted spectra. 
The accuracy of our J-CaPS metallicities is validated against the APOGEE training labels, as shown in Figure~\ref{fig:cv_feh}, which demonstrates good one-to-one correspondence with minimal scatter.
Although the J-CaPS training set contains relatively few stars with [Fe/H] $< -1.5$ and $M < 0.5\,M_\odot$, Figure~\ref{fig:xp_median_spectra} shows that the median XP spectra for these metal-poor, low-mass stars exhibit a metallicity trend consistent with theoretical expectations.
The J-CaPS forward modeling architecture is physically consistent to interpolate across sparsely populated regions of parameter space, allowing it to model spectra that follow physically motivated trends with metallicity across all mass ranges, including the data-sparse regimes encountered in this work.

\section{Mass function results}\label{app:mf}

We provide the inferred stellar mass functions for each metallicity bin in Table~\ref{tab:mf_feh}.
Each table lists the mass function values, $\log_{10}\Phi$ [number\,kpc$^{-3}$\,M$_\odot^{-1}$], as a function of stellar mass derived from our forward modeling approach (Section~\ref{sec:results}).
The mass values represent the centers of each mass bin, and the quoted errors correspond to the $1\sigma$ uncertainties from the posterior distribution of the Poisson model.

\begin{deluxetable}{ccccccccc}
\tablewidth{0pt}
\tablecaption{Stellar mass functions for the metal-poor sample in four metallicity bins \label{tab:mf_feh}}
\tablehead{
\colhead{Mass} & \multicolumn{2}{c}{[Fe/H] = $-2.0$} & \multicolumn{2}{c}{[Fe/H] = $-1.7$} & \multicolumn{2}{c}{[Fe/H] = $-1.5$} & \multicolumn{2}{c}{[Fe/H] = $-1.2$} \\
\colhead{(M$_\odot$)} & \colhead{$\log_{10}(\Phi)$} & \colhead{Error} & \colhead{$\log_{10}(\Phi)$} & \colhead{Error} & \colhead{$\log_{10}(\Phi)$} & \colhead{Error} & \colhead{$\log_{10}(\Phi)$} & \colhead{Error}
}
\startdata
0.18 & 3.07 & 0.35 & 4.16 & 0.20 & 5.59 & 0.04 & 5.40 & 0.05 \\
0.23 & 3.59 & 0.27 & 4.39 & 0.10 & 5.55 & 0.03 & 5.36 & 0.03 \\
0.28 & 3.90 & 0.13 & 4.68 & 0.05 & 5.27 & 0.03 & 5.44 & 0.02 \\
0.34 & 3.86 & 0.10 & 4.51 & 0.05 & 5.03 & 0.03 & 5.32 & 0.02 \\
0.39 & 3.61 & 0.10 & 4.32 & 0.05 & 4.80 & 0.03 & 5.06 & 0.02 \\
0.44 & 3.86 & 0.06 & 4.43 & 0.03 & 4.74 & 0.02 & 4.85 & 0.02 \\
0.50 & 4.08 & 0.03 & 4.47 & 0.02 & 4.66 & 0.02 & 4.66 & 0.02 \\
0.55 & 4.32 & 0.02 & 4.63 & 0.01 & 4.70 & 0.01 & 4.50 & 0.01 \\
0.61 & 4.21 & 0.02 & 4.72 & 0.01 & 4.85 & 0.01 & 4.72 & 0.01 \\
0.66 & 3.44 & 0.04 & 4.26 & 0.02 & 4.70 & 0.01 & 4.82 & 0.01 \\
0.71 & 3.97 & 0.02 & 3.96 & 0.02 & 4.31 & 0.01 & 4.48 & 0.01 \\
0.77 & 3.95 & 0.02 & 3.91 & 0.02 & 4.25 & 0.02 & 4.29 & 0.01 \\
0.82 & 2.54 & 0.11 & 3.22 & 0.05 & 4.20 & 0.02 & 4.41 & 0.01 \\
0.87 & 1.46 & 0.50 & 2.40 & 0.13 & 3.73 & 0.03 & 4.25 & 0.02 \\
\enddata
\tablecomments{Mass function density values are in units of number\,kpc$^{-3}$\,M$_\odot^{-1}$. Errors represent $1\sigma$ uncertainties from the posterior distribution.}
\end{deluxetable}
\bibliography{main}{}
\bibliographystyle{aasjournal}



\end{document}